\documentclass{article}

\usepackage[nonatbib, preprint]{neurips_2024}

\usepackage[utf8]{inputenc} 
\usepackage[T1]{fontenc}    
\usepackage{hyperref}       
\usepackage{url}            
\usepackage{booktabs}       
\usepackage{amsfonts}       
\usepackage{nicefrac}       
\usepackage{microtype}      
\usepackage{xcolor}         
\usepackage{amsmath, amssymb, amsfonts, amsbsy, amsthm, latexsym, epsfig, graphicx, enumitem, bbm, xcolor, caption, subcaption}
\usepackage[style=numeric,sorting=none]{biblatex}

\newtheorem{theorem}{Theorem}

\newtheorem{lemma}[theorem]{Lemma}
\newtheorem{corollary}[theorem]{Corollary}
\theoremstyle{remark}
\newtheorem*{remark}{Remark}
\newcommand{\R}{\mathbb{R}}

\newcommand{\E}{\mathbb{E}}

\DeclareMathOperator{\var}{Var}

\newcommand{\1}{\mathbbm{1}}
\newcommand{\cond}{\, | \,}

\title{Coin-Flipping In The Brain: \\ Statistical Learning with Neuronal Assemblies}

\author{%
  Max Dabagia \\
  School of Computer Science\\
  Georgia Institute of Technology\\
  Atlanta, GA 30332 \\
  \texttt{maxdabagia@gatech.edu} \\
  \And
  Daniel Mitropolsky\\
  Department of Computer Science\\
  Columbia University\\
  New York, NY 10027\\
  \texttt{mitropolsky@cs.columbia.edu}
  \And
  Christos H. Papadimitriou\\
  Department of Computer Science\\
  Columbia University\\
  New York, NY 10027\\
  \texttt{christos@cs.columbia.edu}
  \And
  Santosh S. Vempala \\
  School of Computer Science\\
  Georgia Institute of Technology\\
  Atlanta, GA 30332 \\
  \texttt{vempala@cc.gatech.edu}
}

\begin{document}

\maketitle

\begin{abstract}
How intelligence arises from the brain is a central problem in science. A crucial aspect of intelligence is dealing with uncertainty – developing good predictions about one’s environment, and converting these predictions into decisions. The brain itself seems to be noisy at many levels, from chemical
processes which drive development and neuronal activity to trial variability of responses to stimuli. One hypothesis is that the noise inherent to the brain’s mechanisms is used to sample from a model of the world and generate predictions. To test this hypothesis, we study the emergence of statistical learning in NEMO, a biologically plausible computational model of the brain based on stylized neurons and synapses, plasticity, and inhibition, and giving rise to
assemblies --- a group of neurons whose coordinated firing
is tantamount to recalling a location, concept, memory, or other primitive item of cognition. We show in theory and simulation that connections between assemblies record statistics, and ambient noise can be harnessed to make probabilistic choices between assemblies. This allows NEMO to create internal models such as Markov chains entirely from the presentation of sequences of stimuli. Our results provide a foundation for biologically plausible probabilistic computation, and add theoretical support to the hypothesis that noise is a useful component of the brain’s mechanism for cognition.
\end{abstract}


\section{Introduction}
Any attempt to explain how intelligence arises from the brain must address the organ's remarkable statistical capabilities --- extracting, computing with, and reproducing statistical structure in its environment. A century of work and an enormous body of literature in psychology has quantified these capacities in humans and other animals, across many domains \cite{kelly1994domain, knowlton1994probabilistic, saffran1996word}. Although the mechanisms underlying statistical learning remain poorly understood \cite{batterink2019understanding}, they appear to be subconscious \cite{hasher1984automatic, turk2005automaticity}, and at least partially modality-specific \cite{frost2015domain}, i.e. implemented in the higher-level sensory regions of the brain, rather than via a dedicated mechanism. This raises two questions: What is the most basic level on which statistical learning can be understood to --- for example the neuron, the circuit, or the entire brain? Second, what are the simple, general mechanisms which implement statistical learning? 

In neuroscience, the brain is commonly thought to be inherently noisy at many levels, from the stochasticity of developmental processes \cite{joober2021randomness} and synaptic transmission \cite{branco2009probability,cannon2010stochastic} all the way to the response variability of neurons to identical patterns of stimuli \cite{tolhurst1983statistical, azouz1999cellular}. Various proposals for modeling the computation performed by the brain have postulated that this stochasticity serves a purpose, from improving the magnitude of neural responses \cite{rudolph2001neocortical} to boosting their robustness \cite{anderson2000contribution}. Of particular interest in the context of statistical learning are the many approaches which frame variability as the result, or enabler, of the brain performing \emph{sampling} \cite{hoyer2002interpreting, daw2008pigeon, shi2008performing, levy2008modeling, gershman2009perceptual, buesing2011neural}.

In this work, we propose a stylized neuronal mechanism which leads to learning the statistical properties of stimuli, in the sense that internally-generated randomness leads to sampling from the stimulus distributions. In our model sampling emerges at the level of assemblies: groups of densely-interconnected neurons, which, through their simultaneous firing, encode cognitively-relevant items. Here, the encoded item may be thought of as an outcome of a random variable, and the probability the assembly fires approximates the observed frequency of that event. The ingredients for this mechanism are simple and eminently biologically plausible: random connectivity, inhibition-induced competition, a Hebbian plasticity rule, and noisy neuronal activations are sufficient. Our model is a slight extension to NEMO \cite{papadimitriou2020brain}, a biologically plausible model of the brain in which assemblies of neurons rapidly emerge and general deterministic computation can be carried out by presentation of stimuli~\cite{dabagia2024computation}.

Assembly-level sampling occurs as follows (Theorem \ref{thm:coinflip}): A set of neurons subject to a $k$-winners-take-all dynamic --- NEMO's way of modeling local inhibition --- gives rise to densely-intraconnected assemblies. When these neurons are excited by some external stimulus, the activation of each neuron is perturbed by random noise. Due to inhibition, only a subset of these neurons fire, which in general includes neurons from several assemblies. The random perturbation ensures that one assembly likely enjoys a slight numerical advantage over the others, and the dense intraconnectivity of the assemblies ensures that this initial leader quickly dominates and fires alone. Put differently, assemblies may be viewed as the attractors of the dynamical system, and the effect of randomness is to choose an initial state distributed among these various basins of attraction. In the language of probability, these assemblies represent outcomes of a random variable, while the external stimulus represents a conditional event (e.g. the known value of another random variable). The distribution over outcomes is encoded by the weights from the external stimulus (Corollary \ref{cor:target}), so that an appropriate Hebbian plasticity rule makes it possible to learn the distribution from observing samples (Lemma \ref{lemma:plasticity}).
This is the fundamental building block of our implementation of statistical learning, by which one assembly can be sampled out of a set of possibilities, according to a conditional distribution which depends on external activity and is learned from experience.

Going further, we show in experiment and in theory how this building block naturally leads to the neural implementation of more complex probabilistic models, such as Markov chains (Corollary \ref{cor:markovchain}) and, as a structured extension thereof, a trigram model of language. This generalizes previous work in the NEMO model regarding the memorization of \emph{deterministic} sequences \cite{dabagia2024computation}. The number of assemblies used in these models is proportional to the number of states in the Markov chain, which means that Markov chains can be implemented highly efficiently in terms of neural ``memory". As a demonstration, we show that an assembly-based model can extract and reproduce the lower-order statistics of a small natural language corpus, {\em with no specialized instructions}.

Taken together, our results have several implications. The first is a concrete proposal for how probabilistic computation (and statistical learning more specifically) can be performed by the brain on the level of firing neurons and synapses. This proposal bolsters the hypothesis that assemblies are the fundamental encoding strategy of the brain, as a sparse population code that supports statistical learning. Moreover, we make precise predictions about the functional role of plasticity rules. We also contribute to a line of theoretical work \cite{hoyer2002interpreting, gershman2009perceptual, buesing2011neural} positing a key computational role for noise in the brain.





\section{Background}
\subsection{The Statistical Brain}
A wide body of work in psychology and cognitive science has provided evidence for the sensitivity of humans and other animals to the statistics of their environments. A line of foundational studies \cite{simpson1961psychophysical, erlick1964absolute, pitz1965response, pitz1966sequential} examined the ability of humans to estimate the proportion of different stimuli in a stream presented to them; the overall consensus is that humans are surprisingly good at this task, especially when the classes of stimuli are all relatively common. 
Beyond humans, the tendency of an animal to adjust its foraging strategy such that the amount of time it spends at various feeding sites matches the amount of food or probability of receiving food at those sites is exceptionally well-documented, across clades ranging from fish to passerine birds to rats \cite{brunswik1939probability, estes1964probability, bitterman1965phyletic}. In the study of animal behavior this is known as the ``ideal free distribution'', as the animal has \emph{ideal} (complete) knowledge of the distribution of food and \emph{free} movement. Although counterintuitive, since the optimal strategy is to always choose the most rewarding site, the ideal free distribution is believed to be more stable in a competitive environment \cite{gallistel1990organization}.

Humans and other animals have also demonstrated awareness of higher-order statistics, i.e., the frequency of co-occurrence of stimuli. A landmark study in language learning \cite{saffran1996word} showed that adult humans distinguished between frequently and less-frequently heard pairs of syllables, which was later replicated in infants \cite{saffran1996statistical} and tamarin monkeys \cite{hauser2001segmentation}. Other studies showed similar capabilities in the visual system \cite{fiser2001unsupervised, turk2005automaticity}. Another line of work uses MEG/EEG measures of surprise to gauge subject's responses to sequences of stimuli. Various probabilistic structures have been examined: Furl et al. \cite{furl2011neural} generated the sequence from a Markov chain, while others considered triplets (i.e. trigrams) of stimuli \cite{paraskevopoulos2012statistical, daikoku2014implicit, daikoku2015statistical}. After exposure to the sequence, subjects are reliably more surprised by less-probable transitions.

In contrast to this rich body of behavioral work, the neural basis of statistical learning is poorly understood. Various studies have found that processing of statistical structure is correlated with activity in higher-level, sensory modality-specific areas, in both speech \cite{mcnealy2006cracking, cunillera2009time, karuza2013neural} and vision \cite{turk2009neural, karuza2017neural}. In combination with behavioral studies which show that individual subjects performing statistical learning exhibit different biases \cite{conway2005modality}, uncorrelated performance \cite{siegelman2015statistical}, and the tendency not to generalize statistical structure \cite{conway2006statistical} across different sensory modalities, it has been postulated that statistical learning is implemented by mechanisms local to various cortical regions \cite{frost2015domain}. On the other hand, activity in the brain's general memory systems, especially the hippocampus, is also associated with statistical learning \cite{turk2010implicit, schapiro2012shaping}, although since subjects with damage to these areas still display statistical learning capabilities \cite{covington2018necessity} it is unclear to what extent they are functionally involved in these processes (see Batterink et al. \cite{batterink2019understanding} for a more detailed discussion).


\subsection{Related Work}
NEMO models the brain as a dynamical system arising from a small set of biologically well-grounded mechanisms. NEMO contrasts with Valiant's neuroidal model \cite{valiant2000circuits, valiant2005memorization} in which individual neurons are capable of arbitrary state changes; as well as with 
attempts to implement backpropagation by neural circuitry \cite{lillicrap2020backpropagation}, which rely on hypothetical connectivity and highly-precise coordination; and also with Willshaw and Hopfield network models, which are high-level models of associative memory which deviate from the dynamics and capabilities of the brain. NEMO emphasizes emergent neuronal coordination; assemblies arise to represent external stimuli \cite{papadimitriou2019random} through the interplay of Hebbian plasticity with inhibition-induced competition. Later work in this model has studied other cognitive algorithms that arise as emergent structures of NEMO dynamics, include linear classifiers \cite{dabagia2022assemblies} and memorization of temporal sequences of activations \cite{dabagia2024computation}, while more engineered constructions have been given for natural language \cite{mitropolsky2021biologically, mitropolsky2022center, mitropolsky2023architecture} and implementing finite state and even Turing machines \cite{dabagia2024computation}. Additionally, a related model, with connectivity parameterized by a geometric random graph (as opposed to Erd\"os-Renyi) was shown to display assembly-like behavior even without plasticity \cite{reid2023kcap}.


\paragraph{The brain as a probabilistic model.} The observation that many of the problems the brain solves have a probabilistic or statistical character is integral to cognitive science \cite{rao2002probabilistic, doya2007bayesian} and there have been many attempts to frame computational problems in terms of probabilistic inference \cite{lee2003hierarchical, kersten2004object, kording2004bayesian}. Among neural network models that implement probabilistic inference, the best known are Boltzmann machines \cite{ackley1985learning}, which may be thought of as a probabilistic version of the Hopfield network. The weights of a Boltzmann machine parameterize a restricted class of distributions on $\{0, 1\}^n$, which can be sampled from using Gibbs sampling (a Markov chain Monte Carlo algorithm). These models drew interest in neuroscience since they can be trained using a biologically-plausible synapse-local algorithm, even though the sampling algorithm does not strongly resemble the dynamics of the brain. Later work by Buesing et al. \cite{buesing2011neural} gave a spiking neural network implementation which addresses the latter issue while retaining the biologically-plausible training algorithm. This model inherits two weaknesses from the Boltzmann machine as a model of the brain. The first is that random variables are represented by single spiking units, which requires high reliability of individual neurons and contradicts the widely accepted population coding hypothesis \cite{pouget2000information}. Recent work in has shown the possibility of forming multi-neuron representations of concepts in spiking neural networks \cite{lynch2024multi}, but it remains unclear whether similar ideas apply to the Boltzmann machine implementation. The second is that distributions outside of the class of Boltzmann distributions (i.e. exactly those that are parameterized by the weights of a Boltzmann machine) require deeper networks; how the brain trains its deep networks and what functions it can learn through them remain major open questions.

\section{A Model of the Brain}
In this section we describe NEMO, the mathematical model of the brain we use, along with certain novel extensions, mainly a family of more refined plasticity rules.

\paragraph{Connectivity.} The \emph{brain} is divided into a finite number of \emph{brain areas}, each of which consists of $n$ neurons. Each brain area is recurrently connected by a directed random graph, where each directed edge (or synapse) is present with probability $p$, independently of all others. Each pair of brain areas may be directionally connected, and if so by a directed bipartite random graph with the same edge probability $p$. All synapses have a weight, assumed to be initially $1$. Stimuli are presented in special \emph{input} areas.

\paragraph{The dynamical system.} Time proceeds in discrete steps. At each time step, every neuron in the graph computes its total input, which the is the sum of all weights of incoming synapses from neurons which fired on the previous step. 
In each area, the top $k$ neurons with highest total input fire, and the rest do not. This $k$-winners-take-all process, or $k$-cap\footnote{We will also refer to the set of firing neurons in a brain area as a ``cap''.}, models local inhibition and abstracts excitatory/inhibitory balance. Synaptic weights are nonnegative and subject to Hebbian plasticity: each time neuron $j$ fires followed by neuron $i$, the weight $w_{ij}$ from $j$ to $i$ increases by $\pi(w_{ij})$, where $\pi \colon [0, \infty) \to [0, \infty)$ is the plasticity function. Formally, with $x_A(t)$ the set of neurons firing in $A$, $W_{A,B}$ the weights from area $B$ to area $A$, and $\mathcal{A}$ the set of areas, the update equations are

\begin{align*}
   x_A(t+1) &= k\text{-cap}\left(\sum_{B \in \mathcal A} W_{A, B}(t) x_B(t)\right) \hspace{4.2cm} \forall A \in \mathcal A\\
    W_{A, B}(t+1) &= W_{A, B}(t) + x_{A}(t+1) \cdot x_B(t)^\top \odot \pi(W_{A, B}(t)) \hspace{2cm} \forall A, B \in \mathcal A
\end{align*}
where $\odot$ is element-wise multiplication and the scalar function $\pi$ is applied element-wise\footnote{So that the update rule does not adjust the weight of non-existent synapses, we will always assume that $\pi(0) = 0$.}. The function $k\text{-cap}$ maps a $n$-dimensional vector to the indicator of its $k$ largest elements, with ties broken according to some fixed order, and so $x_A(t)$ is a $\{0, 1\}^n$ vector with either $0$ or $k$ nonzero elements. Neurons may also have their activity perturbed by random noise, in which case the update is
\[x_A(t+1) = k\text{-cap}\left(\sum_{B \in \mathcal A} W_{B, A}(t) x_B(t) + z_A(t)\right)\]
where $z_A(t)$ is a random vector with components independent of each other and the random graph.

In more complex algorithms, we will sometimes assume that brain areas are periodically completely inhibited, or the connections between them are deactivated, especially in patterns of alternation or ``lazy'' alternation -- for instance, area $A$ is allowed to fire once (i.e. form a top $k$ cap) while $B$ is inhibited, then area $B$ is allowed to fire a fixed number of times while $A$ is inhibited, and this process repeats. Since this inhibitory activity does not depend on the state of the model we exclude it from the update equations above for simplicity. In the brain, such functionality might be achieved by long-range interneurons which inhibit one area when excited by another, or as a result of the brain's more global rhythms \cite{buzsaki2010neural}.

\begin{figure}
    \centering
    \includegraphics[width=0.85\textwidth]{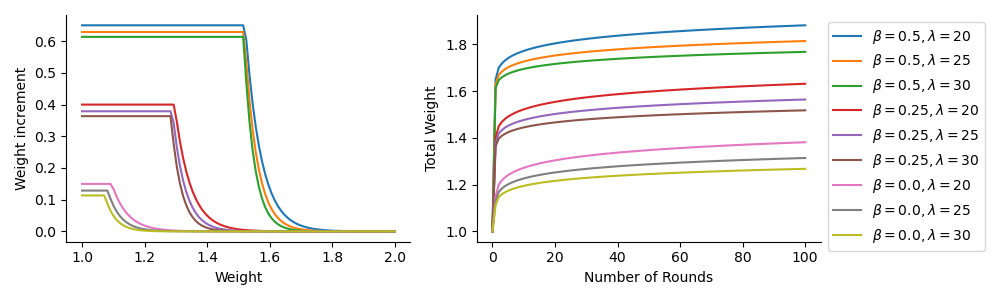}
    \caption{On the left we exhibit the plasticity increment, $\min\{\alpha , e^{\lambda (1 + \beta - w(t))} \}$, as a function of the weight; on the right is the total weight versus the number of rounds of training. Here, $\alpha = \beta + \log \lambda / \lambda$ as we assume in all of our proofs and experiments.}
    \label{fig:plasticity}
\end{figure}

\paragraph{Plasticity rules.} In our experiments and theoretical results we will consider plasticity rules of the form $w(t+1)-w(t)=\pi(w(t)) = \min\{\alpha , e^{\lambda (1 + \beta - w(t))} \}$, where $\alpha, \beta,$ and $\lambda$ are the parameters of the rule.
See Lemma \ref{lemma:plasticity} for a derivation of how this family captures {\em softmax}. Qualitatively, this rule provides a strong increase to the synaptic weight the first few times neurons fire in sequence, which rapidly tapers off with repeated presentations. In Figure \ref{fig:plasticity}, we plot this rule and the resulting weights for various parameter choices. Although it does not impose a hard ceiling on the synaptic weights, this rule bears some similarities to other Hebbian plasticity rules, such as Oja's \cite{oja1982simplified}, remarkable for its ability to perform principle component analysis.

\paragraph{Assemblies.} In this work, we will refer to sets of $k$ neurons in a single brain area as \emph{assemblies} when the internal synaptic weights of the set have been sufficently strengthened. We will also refer to the ``weight" from one assembly to another when we scale all the synapses between them by the same factor. A major result of previous work is that assemblies emerge naturally from the NEMO dynamical system\footnote{In fact, the assemblies obtained this way have additional properties, such as synapse density higher than the base synapse density $p$.}\cite{papadimitriou2020brain}; in this work, we take assemblies as a starting point for representing inputs and outputs to NEMO.



\section{Computational Experiments}
\subsection{The assembly coin-flip}
\begin{figure}[b]
    \centering
    \includegraphics[width=0.7\linewidth]{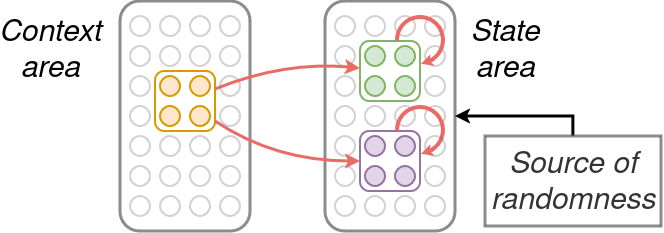}
    \caption{Choosing one of two assemblies to fire, based on context. The probability distribution depends on the weights of each assembly from the context (Theorem \ref{thm:coinflip}).}
    \label{fig:coinflip-schematic}
\end{figure}

We first consider the situation where weights from some input assembly $I$ to assemblies $A_1, \ldots, A_m$ in area $S$ are increased (Figure \ref{fig:coinflip-schematic}). Then $I$ fires, while the activations of neurons in $S$ are perturbed by Gaussian noise; in subsequent rounds, only $S$ fires. A preliminary observation is that if any of $A_1, \ldots, A_m$ has sufficiently large weight from $I$, then after a few rounds of firing the set of firing neurons converges to one of $A_1, \ldots, A_m$. In every experiment in this section, we have $n=25000, k=500, p=0.1$, and the standard deviation of the noise is $5 \sqrt{kp}$.

\begin{figure}
    \centering
    \begin{subfigure}{0.3\textwidth}
    \includegraphics[width=\textwidth]{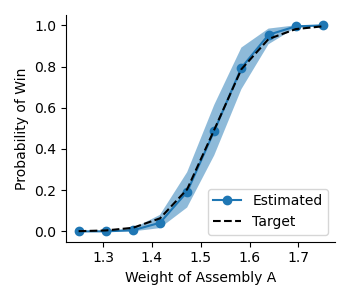}
    \caption{}
    \end{subfigure}
    \hfill
    \begin{subfigure}{0.69\textwidth}
    \includegraphics[width=\textwidth]{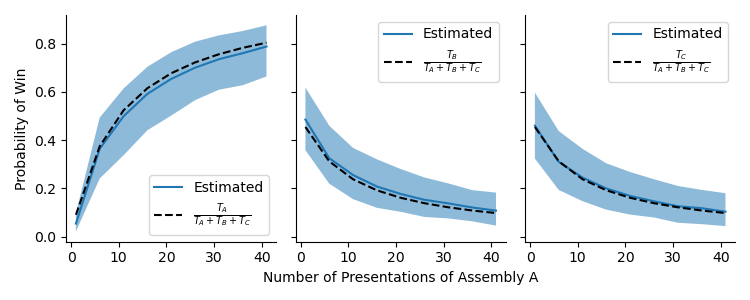}
    \caption{}
    \end{subfigure}
    \caption{Variation in the probability that assembly $A$ wins as its weight from the context assembly is varied, while the weights to $B$ and $C$ are held constant (and equal). On the left the weight is directly adjusted and compared against a best-fitting softmax function (dashed line); on the right all weights are updated incrementally according to a plasticity rule which approximately recovers the observed frequency (dashed line). Dark center line is the mean, while shaded area is the $[0.05, 0.95]$ quantile range, across 100 trials.}
    \label{fig:coinflip-exp}
\end{figure}

In our first experiment (Figure \ref{fig:coinflip-exp}(a)), we fix a sample of the random graph, and measure how the probability that the firing neurons converge to $A_1$ changes as the weight from $I$ to $A_1$ is varied, while weights from $I$ to $A_2$ and $A_3$ are fixed (here at $2$). The internal weights of $A_1, A_2, A_3$ are set to $2$ as well. We repeat the process for $1000$ realizations of the random noise, and compute the empirical frequency that $A_1$ fires. We also plot the softmax function of the weights, where the rate parameter $\lambda$ is chosen empirically to minimize the error between the softmax and the empirical frequency (averaged over graphs). Qualitatively and empirically, this demonstrates that the probability distribution over assemblies is approximately a softmax of the incoming weights to the assemblies (see also Theorem \ref{thm:coinflip}).

In the previous experiment, we directly set the synaptic weights between the input and output assemblies, and studied the resulting probability distribution (of the firing neurons converging to an output assembly). In other words, this shows how \emph{sample} a random variable, assuming its distribution has somehow already been learned and ``stored" in the synaptic weights. 
However, the same neural representations of the outcomes of random variable $A$, namely the assemblies $A_i$, can be used to input \emph{observations} of the random variable during learning. It would be remarkable if, after presenting samples in this way, the probability distribution obtained by running the \emph{previous} experiment (for sampling) yields the empirical distribution seen during training (i.e., the probability the firing neurons converge to $A_i$ is the frequency with which $A_i$ was presented).

This motivates our second experiment (Figure \ref{fig:coinflip-exp}(b)), where we fix a random graph, and then present training samples to the model as follows: first fire $I$, and then $A_i$ repeatedly $T_i$ times. We then measure how the probability that each of $A_1, A_2, A_3$ fire changes as $T_1$ varies, with $T_2 = T_3 = 5$. We also plot the empirical frequency, $\frac{T_i}{T_1 + T_2 + T_3}$. The parameters of the plasticity rule are $\alpha = 0.63$, $\beta = 0.5$, and $\lambda = 26$. Remarkably, plasticity updates suffice to arrive at weights that sample from the empirical distribution! We refer the reader to Corollary \ref{cor:target} and Lemma \ref{lemma:plasticity} for a formal statement of this result and derivation of this plasticity rule.

\begin{figure}[h]
    \centering
    \begin{subfigure}{0.35\linewidth}
        \includegraphics[width=\textwidth]{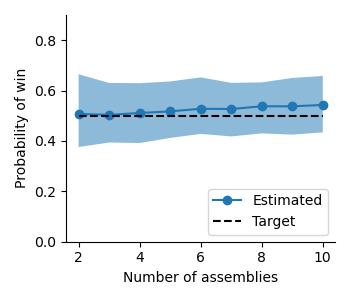}
        \caption{}
    \end{subfigure}
    \begin{subfigure}{0.35\linewidth}
        \includegraphics[width=\textwidth]{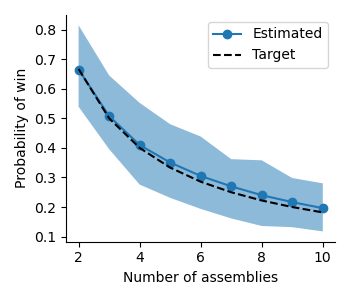}
        \caption{}
    \end{subfigure}
    \caption{Variation in the probability of an assembly winning, as the support of the training distribution grows to include more assemblies, under a plasticity rule that approximately recovers the training distribution. For two different sequences of distributions, we plot the empirical probability of assembly $A_1$ firing (blue dotted line) along with the target probability from the training distribution (black dashed). In (a) and (b) $A_2, \ldots, A_m$ fire $5$ times during training, for each $m=2,\ldots,10$, while in (a) $A_1$ fires $5(m-1)$ times and in (b) $A_1$ fires $10$ times for each $m$.}
    \label{fig:inc-outcomes}
\end{figure}

Finally, to verify that the same plasticity rule suffices to learn distributions over various numbers of assemblies (outcomes), in Figure \ref{fig:inc-outcomes} compare the learned distribution as the number of assemblies that fired during training changes. Notably, a single plasticity rule approximately recovers the observed frequency, even when the number of outcomes is relatively large. In the first experiment (Figure \ref{fig:inc-outcomes}(a)) we vary the number of outcomes, while $A_1$ fires $5(m-1)$ times and $A_2, \ldots, A_m$ fire $5$ times during training; in the second (Figure \ref{fig:inc-outcomes}), $A_1$ fires $10$ times and $A_2, \ldots, A_m$ fire $5$ times. For both, we report the empirical frequency of $A_1$, estimated from $500$ samples over $100$ trials.

\subsection{Scaling} 
\begin{figure}[h]
    \centering
    \includegraphics[width=0.5\textwidth]{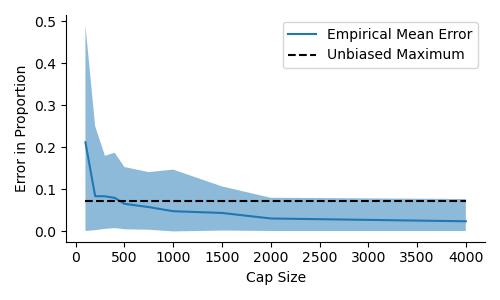}
    \caption{Variation in the error of the learned distribution over two assemblies as the cap size increases. For each of $20$ trials, a new connectivity graph was sampled, the assemblies $A$ and $B$ were made to fire exactly the same number of times, and the resulting distribution was estimated from $500$ samples. The absolute deviation of the frequency with which $A$ won from $1/2$ was recorded for each trial. The dark center line is the mean of this error over trials, the shaded area is the range, and the dashed line is the \emph{expected maximum} if the samples were truly unbiased (i.e. the probability that $A$ comes to fire is exactly $1/2$). }
    \label{fig:inc-size}
\end{figure}

One important observation about the assembly coinflip is that a substantially higher scale is required for the phenomenon to emerge compared to previous work in NEMO \cite{dabagia2022assemblies, dabagia2024computation} -- say, $k=500$ versus $k=30$. Accordingly, we examine this scaling phenomenon quantitatively in Figure \ref{fig:inc-size}. For simplicity we consider the case where the target distribution over assemblies $A$ and $B$ is uniform, so the weights to each are the same -- here, $2$. For each choice of cap size ($k$) we sample $20$ connectivity graphs, increase the weights, and estimate the resulting distribution from $500$ samples. We then plot the resulting absolute deviation of $A$ firing from $1/2$. Notably, the error does not reach $0$, even for very large $k$, but nor do we expect it to, since the maximum deviation of $20$ $(500, 1/2)$ binomial random variables is roughly $0.1$ in expectation (i.e. if the learned distribution were completely unbiased, and the only error was due to sampling).

\subsection{Learning Markov chains}
\begin{figure}
    \centering
    \includegraphics[width=0.9\linewidth]{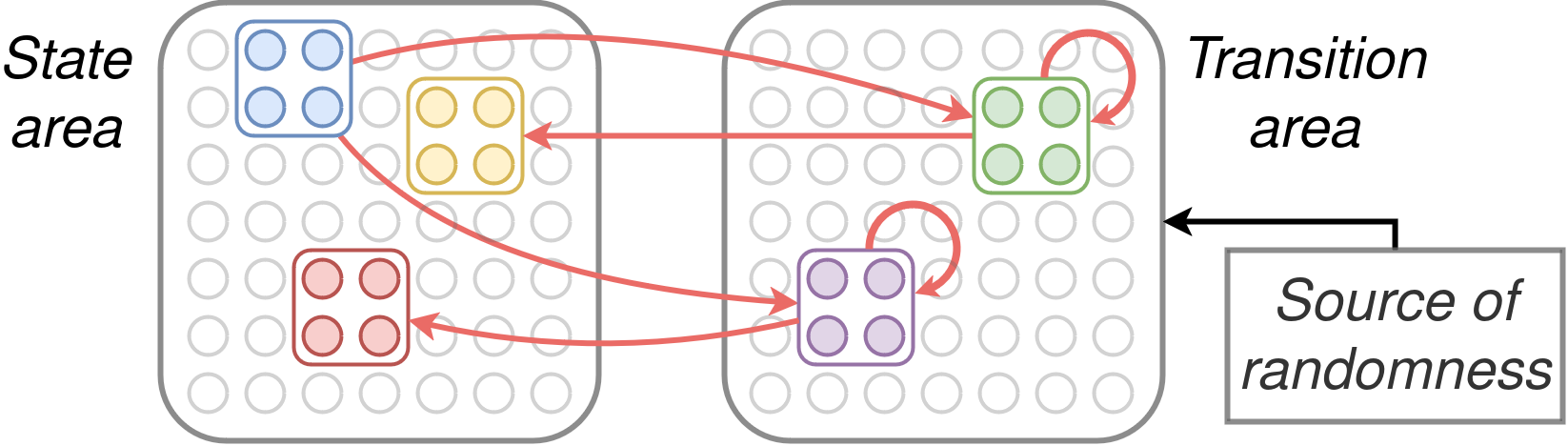}
    \caption{Realizing the transitions of a Markov chain using assemblies (Corollary \ref{cor:markovchain}). Both the state and transition areas contain an assembly for each state. Assemblies in the state area are connected to transition assemblies, with the synaptic weights from a state assembly to the transition assemblies encoding the probability of transitioning to those states.}
    \label{fig:mc-schematic}
\end{figure}

Next, we examine how accurately the transitions of a Markov chain can be learned from a stream of samples, using the basic mechanism demonstrated above (see Corollary \ref{cor:markovchain} for the theoretical formulation of this result). We assume that there ares area $A$ and $B$ which contain an assembly $A_s$, $B_s$ for each state $s$ of the Markov chain (see Figure \ref{fig:mc-schematic}). The areas ``lazily'' alternate firing, so that $A$ fires once while $B$ is inhibited, then $B$ fires $10$ times while $A$ is inhibited, then $A$ fires again once, and so forth.
There are connections in both directions between $A$ and $B$, initially all $1$. Internally, weights within each $A_s$ and $B_{s}$ are strengthened by a factor of $2$.

\begin{figure}[b]
    \centering
    \begin{subfigure}{0.49\textwidth}
        \includegraphics[width=\textwidth]{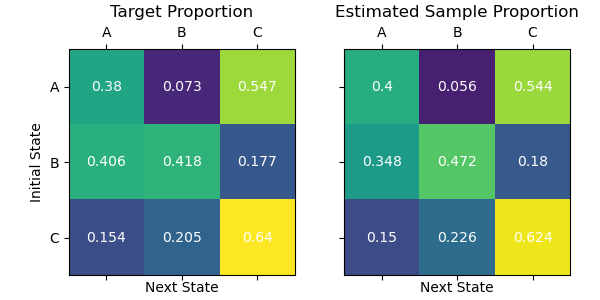}
        \caption{}
    \end{subfigure}
    \hfill
    \begin{subfigure}{0.49\textwidth}
        \includegraphics[width=\textwidth]{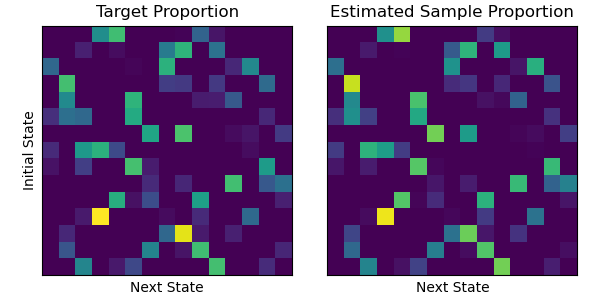}
        \caption{}
    \end{subfigure}
    \caption{We compare an estimate of the transition matrix learned by the model from a stream of samples from a Markov chain with the true transition matrix. This model is trained by presenting sequences of states, sampled from a ground-truth Markov chain, and then firing the appropriate state and transition assemblies in an alternating fashion, with the weights updated by the appropriate Hebbian rule. In both (a) and (b) the left is the ground truth transition matrix (which training samples were generated from) and the right is estimated empirical transition matrix. In (a) we show a chain with a dense transition graph, and in (b) a chain with a sparse transition graph.}
    \label{fig:mc-exp}
\end{figure}

The basic idea of the Markov model is to treat $A$ as the ``context'' area in the coinflip model, and $B$ as the ``state'' area, so that assemblies in $B$ are outcomes with probability conditional on the assembly firing in $A$. The model is trained by presenting a stream of states sampled from the target Markov chain, say $s_1,\ldots,s_t, \ldots$. Training proceeds by firing $A_{s_1}$, followed by $B_{s_2}$, followed by $A_{s_2}$ and so on. In Figure \ref{fig:mc-exp}, the training sequences had length $100 \cdot \# \text{ of states}$. At test time, we sample from the Markov chain by simply firing an initial state, and then observing the activity in area $A$ every $10$ rounds. To estimate the transition probabilties, for each state, we sample $1000$ realizations of random noise and record the empirical frequency of each transition. In Figure \ref{fig:mc-exp} we display the results of learning two chains: one has three states and every transition has positive probability, and the other has $15$ states but each state transitions to only five states with positive probability. For each instance, we display the empirical transition matrix estimated from samples alongside the ground-truth transition matrix from which the training examples were sampled. Here, $n=25000, k=500, p=0.1$, and the noise has standard deviation $5 \sqrt{kp}$, while the plasticity parameters are $\alpha = 0.63$, $\beta = 0.5$, and $\lambda = 26$.

\subsection{Statistical learning of language}
\begin{figure}[t]
    \centering
    \begin{subfigure}{0.4\textwidth}
        \includegraphics[width=\textwidth]{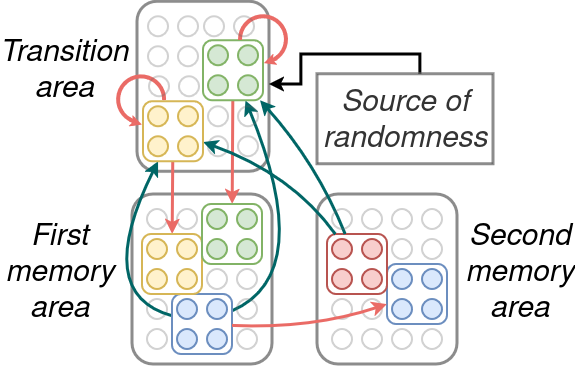}
        \caption{}
    \end{subfigure}
    \hfill
    \begin{subfigure}{0.59\textwidth}
        \centering
        \begin{tabular}{p{0.95\textwidth}}
            \texttt{the owl and the pussycat went to sea in a wood a piggy wig stood with a ring. they took some honey and plenty of money wrapped up in a beautiful pea green boat. they took some honey and plenty of money wrapped up in a wood a piggy wig stood with a ring. they took it away and were married next day by the light of the moon.}
        \end{tabular}
        \caption{}
    \end{subfigure}
    \caption{In (a), we exhibit a schematic of a simple trigram model of language within NEMO. In (b), we show an example string generated after training the model with the text of the poem ``The Owl and the Pussy-Cat'' by Edward Lear, where each word and punctuation mark is a token. The poem contains 109 tokens comprising 182 unique trigrams.}
    \label{fig:trigram}
\end{figure}

Our final experiment applies the probabilistic machinery described so far to construct a primitive model of language (Figure \ref{fig:trigram}). Consider a corpus of language, which is a set of strings comprised of tokens from a lexicon $\mathcal T$. We suppose there exist brain areas $A, B, C$, with connections from $A$ to $B$, $B$ to $A$ and $C$, and $C$ to $A$. The areas contain assemblies $A_\tau, B_\tau, C_\tau$ for each token $\tau \in \mathcal T$; the firing of $A_\tau$ represents token $\tau$ being presented to the model. Internal connections of these assemblies are assumed to be strengthened, and connections from $A_\tau$ to $B_\tau$ and $B_\tau$ to $C_\tau$, but otherwise no weight changes have occurred. Training then proceeds as follows: For a string of tokens $\tau_1, \tau_2, \ldots,$ we fire $A_{\tau_1}, A_{\tau_2}, \ldots$. As this activity propagates to areas $B$ and $C$, the result is strengthening the connections to $A_{\tau_t}$ from $B_{\tau_{t-1}}$ and $C_{\tau_{t-2}}$. Sampling then occurs in area $A$: To sample a third token in the trigram $(\tau_1, \tau_2, \cdot)$, $B_{\tau_2}$ and $C_{\tau_1}$ fire simultaneously, then $A$ is allowed to fire $10$ times, while connections from $A$ to $B$ and from $B$ to $C$ are inhibited, then $A$ is inhibited while inter-areal connections are again enabled. After this process concludes, for some $\tau_3$ we should have $B_{\tau_3}$ along with $C_{\tau_2}$ firing, so sampling may repeat to generate a stream of tokens. Here, we take $n=100,000, k=500, p=0.1$, with plasticity parameters $\alpha=0.63, \beta=0.5, \lambda=60$. The model was trained by presenting the entire corpus $5$ times. We emphasize that no settings or changes were made to NEMO to specialize to the data or output. 

\section{Theoretical Results}

\paragraph{Outline} The presentation of our theoretical results follows the development of the experiments in the previous section. We first establish the basic ingredient of our neuronal-level statistical learner -- randomly choosing one assembly out of a set of possible ``outcome'' assemblies to fire, according to a probability distribution which depends on the weights from the context assembly to the outcome assemblies. This sampling procedure is driven by random noise, which we model here as an i.i.d. perturbation to each neuron's activation. The formal claim that this behavior provably occurs in NEMO (when the set of outcomes has size 2) is the content of Theorem \ref{thm:coinflip}. This gives the distribution as a function of the weights; in Corollary \ref{cor:target}, we determine the weights necessary to achieve a target distribution. Lemma \ref{lemma:plasticity} is the formal claim that the plasticity rule used in our experiments incrementally adjusts the weights during training to ensure that the probability that an assembly fires after training replicates its observed frequency in the training phase. Finally, Corollary \ref{cor:markovchain} uses this mechanism as the building block of a more complex model, namely the simulation of a restricted class of Markov chains using assemblies.

We now proceed with the statement of Theorem \ref{thm:coinflip}. At a high level, this theorem gives sufficient conditions for the parameters of the model such that the distribution of over outcome assemblies is close to a distribution which depends only on the weights of the outcome assemblies (Figure \ref{fig:coinflip-schematic}). It is worth noting that there are two sources of randomness in play: One is the random graph which determines connectivity, and the other arises from the noise model's perturbations to neuronal activations. For our purposes, it is desirable that the realized distribution over outcomes be nearly independent of the random graph. Since complete independence is impossible, we settle for high probability, so that the probability that the realized distribution deviates significantly from a distribution genuinely independent of the connectivity goes to zero as $n$ (the number of neurons) goes to $\infty$.

\begin{theorem}[\textbf{Conditional Coin-Flipping}] \label{thm:coinflip}
    Consider NEMO brain areas $I$ and $M$, with a context set $I_0 \subseteq I$ and outcome sets $A, B \subseteq M$, all disjoint and of size $k$. Let the connections from $I_0$ to $A$ (resp. $B$) be strengthened by a factor of $w_A$ (resp. $w_B$), and the connections from $A$ to $A$ and from $B$ to $B$ be strengthened by a factor of $w_R$, and all other connections from $I_0 \cup A \cup B$ to $M$ have weight $1$. Suppose that $I_0$ fires once, and the total inputs to neurons in $M$ are perturbed by i.i.d. $\mathcal N(\mu, \sigma^2 kp)$ random variables. 
    
    Let $C_t$ denote the set of $k$ neurons firing in $M$ after $t$ rounds. Then there exist $p, w_0, w_R, \sigma, \tau, t_0$ such that if $w_A, w_B \ge w_0$, for all $t \ge t_0$ we have
    \begin{align*}
        \left|\Pr(C_t = A | \mathcal G) - \Phi\left( \frac{w_A - w_B}{\tau}\right) \right| = o(1),\hspace{1cm} \left|\Pr(C_t = B | \mathcal G) - \Phi\left(\frac{w_B - w_A}{\tau}\right) \right| &= o(1)
    \end{align*} with high probability over the random graph $\mathcal G$ as $n \to \infty$. Here, $\Phi$ is the Gaussian CDF\footnote{It is worth noting that the appearance of the Gaussian CDF here is \emph{not} due to the normal distribution of the noise, but rather because under fairly general assumptions sample quantiles are asymptotically normally distributed.}.
\end{theorem}

The upshot of Theorem \ref{thm:coinflip} is that there is an explicit function of the weights only, which does not depend on the actual realization of the underlying graph (only on the parameters of the model), so that the actual probability distribution over outcomes is very close to this function.
A natural question is whether we can set the weights in such a way that some target probability distribution is realized. 
We first characterize in the next corollary the weights necessary to achieve a certain probability distribution on the eventual firing of $A$ or $B$.

\begin{corollary} \label{cor:target}
    Under the assumptions of Theorem \ref{thm:coinflip}, there exist $c, \lambda > 0$ such that, if 
    \begin{align*}
        \left|w_A - \frac{\log c T_A}{\lambda}\right| \le \frac{1}{50\lambda}, \hspace{2cm}
        \left|w_B - \frac{\log c T_B}{\lambda}\right| \le \frac{1}{50\lambda}
    \end{align*}
    then for all $t \ge t_0$ we have
    \[\left|\Pr(C_t = A | \mathcal G) - \frac{T_A}{T_A + T_B} \right| < 1/ 25 + o(1)\]
\end{corollary}
The essence of this result is just showing that the Gaussian CDF $\Phi(x)$ is close to the logistic function ($2$-dimensional softmax), $\frac{e^{\lambda x}}{1 + e^{\lambda x}}$, for appropriate choice of $\lambda$.

We now establish a plasticity rule which, through the dynamics of NEMO, induces weights through training such that the induced probability distribution approximates the observed frequency.


\begin{lemma} \label{lemma:plasticity}
    Under the assumptions of Theorem \ref{thm:coinflip}, suppose that the weight update is
    \[w_{ij}(t+1) = w_{ij}(t) + \max\{\alpha, e^{\lambda (1 + \beta - w_{ij}(t))}\}\]
    when neuron $j$ fired on round $t$ and neuron $i$ fired on round $t+1$, and $w_{ij}(t)$ is the weight of the synapse from $j$ to $i$ on round $t$. Suppose that $A$ fires after $I_0$ $T_A$ times, and $B$ fires after $I_0$ $T_B$ times. Then there exists $T_0, \alpha, \beta, \lambda$ such that for $T_A, T_B \ge T_0$,
    \begin{align*}
        \left|\Pr(C_t = A \cond \mathcal G) - \frac{T_A}{T_A + T_B}\right| < \frac{1}{25} + o(1).
    \end{align*}
\end{lemma}
Note that the plasticity rule is still Hebbian, and completely synapse-local; what is specified is only the magnitude of the change in weight, as a function of the current weight.  This rule is a natural approximation of the ``target'' weights for proportional sampling given in Corollary \ref{cor:target}, as follows. For $t=0$, we have $w(0) = 1$ and the target $w(1) = \lambda^{-1} \log c$, so the incremental change is simply $w(1) - w(0) = \lambda^{-1} \log c$. For $t \ge 1$, the increment is 
\[w(t+1) - w(t) = \frac{\log(c(t+1)) - \log (ct)}{\lambda} = \frac{\log(1+1/t)}{\lambda}.\]
 We use the approximation $\log(1 + 1/t) \approx 1/t$, and then use the assumption that $t = e^{\lambda w(t)} / c$ (which is exactly true for the target weight) to get $w(t+1) - w(t) = \frac{c}{\lambda} e^{-\lambda w(t)}$. We take $c = \lambda e^{\lambda(1 + \beta)}$ and set $\alpha = \beta + \lambda^{-1} \log( \lambda)$ to recover the parameterization of the rule in the lemma.

Using the above result, we show rigorously that a finite, stationary Markov chain with out-degree at most 2 can be simulated in NEMO.

\begin{corollary}[Markov Chain] \label{cor:markovchain} Consider a Markov chain with states $s =1,\ldots,m$, such that the transition probabilities from $s$ to state $\rho(s, i)$ is $p_{s, i}\ge0$ for $i=1,2$ and $0$ for all other states. Consider distinct brain areas $A$ and $B$, with disjoint state sets $A_s \subseteq A, B_s \subseteq B$ for $s=1, \ldots, m$. Suppose that when $A$ fires $B$ fires for $t$ rounds while $A$ is inhibited, then $B$ is inhibited and $A$ is permitted to fire. Furthermore, suppose for each $s=1,\ldots, m$, $i=1,2$, weights from $A_s$ to $B_{\rho(s, i)}$ are strengthened by a factor of $\frac{1}{\lambda}\log(c \cdot p_{s, i}) + 1$, and weights from $B_{s}$ to $A_{s} \cup B_s$ are strengthened by $w_R$. Then there exist $p, c, \lambda, \sigma, t$ such that for all $s=1,\ldots,m$, $i=1,2$, if $A_s$ fires, the probability that $A_{\rho(s, i)}$ fires $t+1$ rounds later deviates from $p_{s, i}$ by no more than $1/50$, w.h.p. as $n \to \infty$.
\end{corollary}
In the notation of the corollary, this means that a model configured as above approximates the Markov chain with states $\{1, \ldots, m\}$, and with transition probability from state $s$ to state $\rho(s, i)$ of $p_{s, i}$. Using the plasticity rule in Lemma \ref{lemma:plasticity}, the transition probabilities can be learned from a sequence $s_1, s_2, \ldots$ sampled from the target Markov chain by firing the assemblies $A_{s_1}, B_{s_2}, A_{s_2}, B_{s_2}, \ldots$. Furthermore, out-degree 2 Markov chains are general in the following sense:

\begin{remark}
Consider a Markov chain with states $[m]$ and transition probability $p_{i, j}$ from state $i$ to state $j$. Then there exists a Markov chain $(X_t)$ with states $[m] \times [m-1]$ and out-degree $2$ such that
\[\Pr(X_{t + \max\{j, m-1\}} = j \cond X_t = i) = p_{i, j}.\]
\end{remark}
Intuitively, this construction reduces a single random choice of one item out of many to a sequence of binary choices, whether to choose a particular item or discard it and repeat with the smaller set. In terms of the transition graph of a Markov chain, it replaces each random transition with a binary tree of transitions, whose leaf vertices correspond to states in the original chain.



\paragraph{The noise assumption.}
In our experiments and proofs, we have assumed that neuronal activations are perturbed by i.i.d. Gaussian noise. Due to the Central Limit Theorem, this is a reasonable way to abstractly model the contribution of many small independent sources of uncertainty. In particular, one can explicitly model the source of randomness in the setting of Theorem \ref{thm:coinflip} by adding an additional area $R$ of size $\sigma^2 n$ with connections to area $A$, and assume that the only randomness is that a \emph{random set} of $k$ neurons fires in $R$. The resulting perturbations to the activations of neurons in $A$ will be independent, identically-distributed binomial $(\sigma^2 n, p)$ random variables, which under the CLT converge in distribution to $\mathcal N(\sigma^2 np, \sigma^2 np(1-p))$ with appropriate normalization.

\section{Proofs}
The proof of Theorem \ref{thm:coinflip} consists of showing that $A$ obtains a decent advantage on the first round using Lemma \ref{lemma:first-round} with the specified probability that $A$ wins altogether, and then showing that this initial lead will quickly grow to activate the entire assembly using Lemma \ref{lemma:conv-w-advantage}.
\begin{proof}[Proof of Theorem \ref{thm:coinflip}]
    First, WLOG, assume that $\mu = 0$. Let $X_i(0)$ denote the input to neuron $i$ when $I_0$ fires, and $Y_i \sim \mathcal N(0, \sigma^2 kp)$ denote the random perturbations; then we have
    \[X_i = \begin{cases} w_A \cdot e(I_0, i) + Y_i & i \in A\\
    w_B \cdot e(I_0, i) + Y_i & i \in B\\
    e(I_0, i) + Y_i & \text{o.w.}
    \end{cases}\]
    We will first establish that $C_t \subseteq A \cup B$ for all $t \ge 0$ w.h.p. by induction. For the base case, note that w.h.p. for all $i \not\in A \cup B$, we have $e(I_0, i) \le kp + \sqrt{3kp \log n}$ by the Chernoff bound; likewise, $Y_i \le \sigma \sqrt{3kp\log n}$. Hence,
    \[X_i \le kp + \sigma \sqrt{3 kp \log n} + o(kp).\]
    On the other hand, for all $i \in A \cup B$, we have
    \[X_i \ge \min\{w_A, w_B\} (1-o(1)) kp - \sigma \sqrt{3 kp \log n} \ge kp + \sigma \sqrt{3 kp \log n} + o(1) \ge \max_{j \not\in A \cup B}X_j\]
    for all sufficiently large $n, \sigma$. Hence, w.h.p. $C_0 \subseteq A \cup B$. Now, suppose that $C_t \subseteq A \cup B$. WLOG, suppose that $|C_t \cap A| \ge k/2$. Then for $i \in A$, we have
    \[X_i(t+1) \ge (1 + w) kp / 2 - o(1) \ge kp + o(1) = \max_{j \not\in A \cup B} X_j(t+1)\]
    for all sufficiently large $n$ w.h.p. Hence, every neuron in $A$ has larger input than every neuron outside of $A \cup B$, and so $C_{t+1} \subseteq A \cup B$. By the principle of induction, it follows that $C_t \subseteq A \cup B$ for all $t \ge 0$.

    Now, let $T_0$ be the median of $\{X_i(0) \colon i \in A \cup B\}$, and note that $i \in C_0$ if and only if $X_i(0) \ge T_0$. By Lemma \ref{lemma:first-round}, we have
    \begin{align*}
        \Pr(|C_0 \cap A| &\ge k/2 + \sqrt{k / \log k} \cond \mathcal G) \ge \Phi\left(\lambda \frac{k\sqrt{p}}{\sigma}(w_A - w_B)\right)\\
        \Pr(|C_0 \cap B| &\ge k/2 + \sqrt{k / \log k} \cond \mathcal G) \ge \Phi\left(\lambda \frac{k\sqrt{p}}{\sigma}(w_B - w_A)\right)
    \end{align*} Now, suppose that $ k - |C_t \cap A| \le \sqrt{k / \log k}$. Since $\sqrt{k / \log k} \ge 6 p^{-1}$,  by Lemma \ref{lemma:conv-w-advantage}, we will have 
    \[|C_{t+1} \cap A| \ge k/2 + \max\{k/2, \frac{\sqrt{kp}}{6}  (|C_t \cap A| - k/2)\}\]
    w.p. at least $1 - \exp\left(-(|C_t \cap A| - k/2)^2 p\right)$.
    Thus, if $|C_0 \cap A| \ge k/2 + \sqrt{k / \log k}$, then for all $t \ge 1$ we have
    \[|C_t \cap A| \ge k/2 + \max\{k/2, \left(\sqrt{kp}{6} \right)^t \sqrt{k / \log k}\}\]
    w.p. $1- o(1)$, and so after $t_0 = O(\log k / \log (kp))$ rounds, we must have $|C_{t_0} \cap A| = k$ w.h.p. A similar argument holds for $B$.
\end{proof}

\begin{proof}[Proof Sketch of Corollary \ref{cor:target}]
    One can verify (even numerically) that
    \[\sup_{x \in \R}\left|\frac{e^{1.6x}}{1 + e^{1.6x}} - \Phi(x)\right| < \frac{1}{50}.\]
    In particular, for $x = \frac{(w_A-w_B)}{\tau}$, and setting $\lambda = 1.6 / \tau$, we have
    \[ \frac{e^{1.6x}}{1+e^{1.6x}} = \frac{e^{1.6 w_A / \tau}}{e^{1.6 w_A / \tau}+ e^{1.6 w_B / \tau}} = \frac{e^{\lambda w_A}}{e^{\lambda w_A} + e^{\lambda w_B}} \]
    Now, for settings of $w_A$ and $w_B$ that satisfy the corollary conditions,
    \[\frac{e^{\lambda w_A}}{e^{\lambda w_A} + e^{\lambda w_B}} \approx \frac{c T_A}{cT_A + cT_B} \approx \frac{T_A}{T_A + T_B}\]
    where $\approx$ denotes an absolute difference of $o(1)$, and so
    \[\left|\Phi\left(\frac{w_A - w_B}{\tau}\right) - \frac{T_A}{T_A + T_B}\right| < 1/ 50 + o(1).\]
    We then take $c$ and $T_0$ large enough to ensure the condition on $w_0$ in Theorem \ref{thm:coinflip} is satisfied.
\end{proof}

\begin{proof}[Proof Sketch of Lemma \ref{lemma:plasticity}]
    We first note that for $\alpha = \beta + \lambda^{-1} \log(\lambda)$, one has $w(1) = \lambda^{-1}\log(c \cdot 1)$ for $c = \lambda e^{\lambda(1 + \beta)}$. 
    The substance of the proof is showing that for all $t \ge 2$, we have
    \[\frac{\log (c t)}{\lambda} \le w(t) \le \frac{\log (c(t + \log t))}{\lambda}.\]
    It is readily verified that $w(t)$ satisfying the above also meets the conditions of Corollary \ref{cor:target} and hence the desired probabilities obtain.
    The above claim reduces to showing that the recurrence relation
    \[x_{n+1} = x_n + e^{-x_n} \hspace{2cm} x_1 = 0\]
    satisfies $\log n \le x_n \le \log(n + \log n)$. This can be verified by induction; clearly it holds for $n=1$. For $n \ge 2$, we have the lower bound
    \[x_{n+1} = x_n + e^{-x_n} \ge \log n + \frac{1}{n} \ge \log (n+1).\]
    For the upper bound, it can be verified that the function 
    \[f(x) = \log(x + \log x) + \frac{1}{x + \log x} - \log(x+1 + \log(x+1))\]
    has a single extremum, at which it is positive, and since $f(1) = 0$ and $\lim_{x \to \infty}f(x) = 0$ the inequality $x_{n+1} \le \log(n+1 + \log(n+1))$ holds.
\end{proof}

\begin{proof}[Proof Sketch of Corollary \ref{cor:markovchain}]
    For each $s=1,\ldots,m$, Corollary \ref{cor:target} asserts that this choice of weights will ensure that when $A_s$ fires, the probability that $B_{\rho(s, i)}$ fires will be $p_{s, i} \pm 1/50$. For sufficiently large $w_R$, when $B_{\rho(s, i)}$ fires $A_{\rho(s, i)}$ will fire immediately after. So, by choosing $t$ larger than $t_0$ in Theorem \ref{thm:coinflip} we can ensure that the entirety of $B_{\rho(s, i)}$ fires when $A$ becomes disinhibited, so the probability that $A_{\rho(s, i)}$ fires in area $A$ $t$ rounds after $A_s$ is very nearly $p_{s, i}$.  
\end{proof}

\begin{lemma} \label{lemma:conv-w-advantage}   For $m \in [n]$, let $X_1, \ldots, X_n$ be i.i.d. $(m, p)$ binomial random variables, with $X_{(i)}$ their $i$th order statistic. Let $\Delta = |m - n/2|$ and set $k = \max\{\lfloor n/2 + \frac{\sqrt{np}}{6} \cdot \Delta \rfloor, n\}$. Then the following hold:
    \begin{enumerate}
        \item For $m \ge n /2 + 6p^{-1}$, we have
        \[\Pr(X_{(n-k)} \le np / 2) \le e^{-\Omega(\Delta^2 p)}\]
        \item For $m \le n /2 - 6p^{-1}$, we have
        \[\Pr(X_{(k)} \ge np / 2) \le e^{-\Omega(\Delta^2 p)}\]
    \end{enumerate}    
\end{lemma}

\begin{proof}
    First, if $\Delta \ge \sqrt{4 n \log n}$, then the Chernoff bound gives
    \[\Pr(X_1 \le np / 2) \le \exp\left(-\frac{4 \log n}{1 + 2 \epsilon}\right) \le \frac{1}{n^2}\]
    and so by the union bound we have $X_{(1)} > np/2$ w.p. $1 - 1/n$. For $\Delta < \sqrt{4n \log n}$,
    set
    \[Y = \sum_{i=1}^n \1_{\{X_i > np / 2\}}\]
    and note that $Y$ is a $(n, q_m)$-binomial random variable, where
    \[q_m = \Pr(X_1 > np / 2)\]
    Moreover,
    \[\Pr(X_{(n-k)} \le np / 2) = \Pr(Y \le k) \le \exp\left(-\frac{(n q_m - k)^2}{2n q_m}\right)\]
    via the Chernoff bound. Then using the Berry-Esseen theorem,
    \[1 - q_m = \Pr(X_1 \le np / 2) \le \Phi\left( -\frac{\Delta p}{\sqrt{mp(1-p)}}\right) + \frac{1}{2\sqrt{m p (1-p)}}\]
    where $\Phi \colon \R \to [0, 1]$ is the normal CDF. By Taylor's theorem, there exists $t \in [0, \Delta p]$ such that
    \[\Phi\left( -\frac{\Delta p}{\sqrt{mp(1-p)}}\right) = \Phi(0) - \frac{\Delta p}{\sqrt{mp(1-p)}} \cdot \phi \left(\frac{t}{\sqrt{mp(1-p)}}\right)\]
    where $\phi \colon \R \to \R$ is the normal PDF.
    Noting that $\phi$ is monotone decreasing for $t \ge 0$, we have
    \[\Phi\left( -\frac{\Delta p}{\sqrt{mp(1-p)}}\right) \le \frac{1}{2} - \frac{\Delta p}{\sqrt{mp(1-p)}} \cdot \frac{\exp\left(- \frac{(\Delta p)^2}{2 mp(1-p)}\right)}{\sqrt{2\pi}}\]
    As $\Delta = o(n)$, $m \le n$, and $\Delta p \ge 6 \ge 2 \sqrt{2 \pi}$, it follows that
    \[q_m \ge \frac{1}{2} + (1 - o(1)) \frac{\Delta p}{\sqrt{8 \pi np}}\]
    Hence,
    \begin{align*}
        n q_m - k \ge \frac{n}{2} + (1 - o(1)) \sqrt{\frac{np}{8 \pi}} \Delta - \frac{n}{2} - \frac{\sqrt{np}}{6} \Delta \ge \frac{\sqrt{np}}{2} \Delta
    \end{align*}
    which gives the claim:
    \begin{align*}
        \Pr(X_{(n-k)} \le np / 2) &\le \exp\left(-\frac{\Delta^2 \cdot np}{8 n q_m}\right) \le \exp\left(-\frac{\Delta^2 p}{4}\right)
    \end{align*}
    The proof of the second claim is similar.
\end{proof}

\begin{lemma} \label{lemma:first-round}
    Let $X_1, X_1', \ldots, X_n, X_n'$, $Y_1, Y_1', \ldots, Y_n, Y_n'$ be independent, with $X_i, X_i' \sim \mathcal B(n, p)$ and $Y_i, Y_i' \sim \mathcal N(0, \sigma^2 np(1-p))$ for all $i=1,\ldots, n$. Let $M$ denote the median of the combined sample $w X_1 + Y_1, w' X_1' + Y_1', \ldots, wX_n + Y_n, w'X_n' + Y_n'$. Then w.h.p. as $n, \sigma \to \infty$, we have 
    \begin{align*}
        \Pr\left(\#\{i=1,\ldots, n\colon wX_i + Y_i \ge M\} \ge \frac{n}{2} + \sqrt{\frac{n}{\log n}} \,\bigg|\, X_1, X_1' \ldots, X_n, X_n'\right)\\
        \ge \Phi\left(\lambda \frac{n\sqrt{p}}{\sigma} (w - w')\right)  - o(1)
    \end{align*} for some constant $\lambda > 0$.
\end{lemma}
\begin{proof}
    Set $Z_i = wX_i + Y_i$ and $Z_i' = w'X_i' + Y_i'$, and let $Z_{(k)}, Z_{(k)}'$ denote their $k$th order statistics, respectively. Note that $\#\{i=1,\ldots,n \colon Z_i \ge M\} \ge k$ if and only if $Z_{(n-k+1)} \ge Z_{(k)}'$. Let $Z_{(k)}(x) = Z_{(k)} \cond \{X = x\}$, and note that $Z_{(k)}(x)$ is the $k$th order statistic of the normal random variables $Y_1 + w x_1, \ldots, Y_n + w x_n$, where $Y_i + wx_i$ has mean $wx_i$ and variance $\sigma^2 np(1-p)$.
    Let $Q(x), R(x)$ be chosen so that
    \begin{align*}
        \frac{1}{n} \sum_{i=1}^n \Phi\left(\frac{Q(x) - wx_i}{\sigma \sqrt{np(1-p)}}\right) &= \frac{1}{2} - \frac{1}{\sqrt{\log n}}\\
        \frac{1}{n} \sum_{i=1}^n \Phi\left(\frac{R(x) - w' x_i}{\sigma \sqrt{np(1-p)}}\right) &= \frac{1}{2} + \frac{1}{\sqrt{\log n}}
    \end{align*}
    noting that this is possible as $\Phi$ is continuous. 
    Denote by $\Sigma(x), \Sigma'(x)$ the quantities
    \begin{align*}
        \Sigma(x) &= \sqrt{\sigma np(1-p)} \frac{\sqrt{\sum_{i=1}^n \Phi\left(\frac{Q(x) - wx_i}{\sigma \sqrt{np(1-p)}}\right)\left(1-\Phi\left(\frac{Q(x) - wx_i}{\sigma \sqrt{np(1-p)}}\right)\right)}}{\sum_{i=1}^n \phi\left(\frac{Q(x) - wx_i}{\sigma \sqrt{np(1-p)}}\right)}\\
        \Sigma'(x) &= \sqrt{\sigma np(1-p)} \frac{\sqrt{\sum_{i=1}^n \Phi\left(\frac{R(x) - w' x_i}{\sigma \sqrt{np(1-p)}}\right)\left(1-\Phi\left(\frac{R(x) - w' x_i}{\sigma \sqrt{np(1-p)}}\right)\right)}}{\sum_{i=1}^n \phi\left(\frac{R(x) - w' x_i}{\sigma \sqrt{np(1-p)}}\right)}
    \end{align*}
    and let $S(x) = Z_{n/2 - \sqrt{n / \log n}}(x), S'(x') = Z_{n/2+\sqrt{n/\log n}}'(x')$.
    By Lemma \ref{lemma:order} (specifically Remark \ref{remark:order}), we have that the random variables
    \[\frac{S(X) - Q(X)}{\Sigma(X)}, \frac{S'(X') - R(X')}{\Sigma'(X')}\]
    converge in distribution to $\mathcal N(0, 1)$ almost surely (over $X$ and $X')$. Let $x$, $x'$ be chosen so that this holds. As $S(x), S'(x')$ are independent, the random variable
    \[\frac{Z_{(n/2-\Delta)}(x) - Z_{(n/2+\Delta)}'(x') - (Q(x) - R(x'))}{\sqrt{\Sigma(x)^2 + \Sigma'(x')^2}}\]
    also converges to $\mathcal N(0, 1)$ in distribution.
    Thus, for $W \sim \mathcal N(0, 1)$, 
    \[\Pr(S(x) - S'(x') \ge 0) \ge \Pr\left(W \ge \frac{R(x') - Q(x)}{\sqrt{\Sigma(x)^2 + \Sigma(x')^2}}\right) - o(1)\]
    Now, note that if $Q(x) \le t$, then by monotonicity
    \[\sum_{i=1}^n \Phi\left(\frac{t - w x_i}{\sigma \sqrt{np(1-p)}}\right) \ge \frac{n}{2} - \sqrt{\frac{n}{\log n}}.\]
    Let $t_q$ be chosen so that $\E \Phi\left(\frac{t_q - w X_1}{\sigma\sqrt{np(1-p)}}\right) = q$. As $\Phi$ is $1/\sqrt{2\pi}$-Lipschitz, by Lemma \ref{lemma:varfn} we have
    \[\var \Phi\left(\frac{t_q-w X_1}{\sigma \sqrt{np(1-p)}}\right) \le \frac{w^2}{2\pi} \frac{\var X_1}{\sigma^2 np(1-p)} = \frac{w^2}{2\pi \sigma^2}.\] 
    Furthermore, clearly
    \[\E \left|\Phi\left(\frac{t_q-w X_1}{\sigma \sqrt{np(1-p)}}\right) - q\right|^3 \le 1.\]
    Set $q = 1/2 - \frac{1}{\sqrt{n \log n}} - \frac{1}{\sqrt{\sigma n}}$. The Berry-Esseen theorem then provides that
    \begin{align*}
        \Pr(Q(X) \le t_q) &\le \Pr\left(\sum_{i=1}^n \Phi\left(\frac{t_q-wX_i}{\sigma \sqrt{np(1-p)}}\right) \ge n/2 - \sqrt{n/\log n}\right)\\
        &\le \Phi\left( -\frac{n/2 - \sqrt{n/\log n} - (n/2 - \sqrt{n / \log n} - \sqrt{n / \sigma})}{w \sqrt{n / 2\pi \sigma^2}}\right) + o(1)\\
        &= \Phi\left(-\sqrt{2 \pi \sigma} / w\right) + o(1)\\
        &= \Phi(-\omega(1)) + o(1) = o(1).
    \end{align*}
    A similar argument establishes that $R(X') \le t_{1-q}$ w.p. $1 - o(1)$. If the above both hold, then
    \[R(X') - Q(X) \le t_{1-q} - t_{q} \le (w' - w)np + \sigma \sqrt{np(1-p)} \cdot o(n^{-1/2})\]
    Now, note that
    \[\E \left[\Phi(\frac{t_q-wX_1}{\sigma \sqrt{np(1-p)}}) \left(1-\Phi(\frac{t_q-X_1}{\sigma \sqrt{np(1-p)}})\right)\right] = q(1-q) - \var \Phi\left(\frac{t_q-wX_1}{\sigma \sqrt{np(1-p)}}\right)\]
    Hence,
    \[\E \left[\Phi(\frac{t_q-wX_1}{\sigma \sqrt{np(1-p)}}) \left(1-\Phi(\frac{t_q-wX_1}{\sigma \sqrt{np(1-p)}})\right)\right] \ge q(1-q) - \frac{w^2}{2\pi \sigma^2}.\]
    By Hoeffding's inequality,
    \begin{align*}
        \sum_{i=1}^n \Phi(\frac{t_q-X_i}{\sigma \sqrt{np(1-p)}}) \left(1-\Phi(\frac{t_q-X_i}{\sigma \sqrt{np(1-p)}})\right) \ge (1-o(1))n\left(\frac{1}{4} - \frac{w^2}{2\pi \sigma^2}\right) = (1 - o(1)) \frac{n}{4}
    \end{align*} with high probability. On the other hand, clearly
    \[\sum_{i=1}^n \phi\left(\frac{Q(x) - wx_i}{\sigma \sqrt{np(1-p)}}\right) \le n\]
    and so
    \begin{align*}
        \sqrt{\Sigma(X)^2 + \Sigma'(X')^2} &= \sigma \sqrt{np(1-p)} \sqrt{2\frac{(1-o(1))n/4}{n^2}} = (1-o(1)) \sigma \sqrt{ np(1-p) / 2n}
    \end{align*}
    again w.h.p.
    Finally, with $x, x'$ satisfying all of the above conditions, we have
    \begin{align*}
        \Pr(S(x) \ge S'(x')) &\ge \Pr\left(N \ge \frac{(w' - w)np + \sigma \sqrt{np(1-p)} \cdot o(n^{-1/2})}{\sigma \sqrt{np(1-p) / 2n}} \right) - o(1)\\
        &= \Pr(N \ge \lambda \frac{n\sqrt{p}}{\sigma} (w' - w) + o(1)) - o(1)\\
        &= \Phi\left(\lambda \frac{n\sqrt{p}}{\sigma} (w - w')\right) - o(1).
    \end{align*}
    Since these conditions hold w.h.p. for $X$ and $X'$, the proof is complete.
\end{proof}

\begin{lemma} \label{lemma:order}
    Let $X_1,\ldots, X_n$ be independent but not necessarily identically distributed random variables, where $X_i$ has continuous p.d.f. and c.d.f. $f_i, F_i$, respectively. Let $X_{(k)}$ denote their $k$th order statistic, and $\overline f_n = \frac{1}{n} \sum_{i=1}^n f_i$, $\overline F_n = \frac{1}{n} \sum_{i=1}^n F_i$. Suppose that the following conditions hold:
    \begin{enumerate}[label=(\roman*)]
        \item $\lim_{n \to\infty} k / n \in (0, 1)$;
        \item There exists a convergent sequence $(t_n)_n$ such that $|\overline F_n(t_n) - k/n| = o(n^{-1/2})$;
        \item With $\sigma_n^2 = \frac{1}{n}\sum_{i=1}^n F_i(t_n) (1 - F_i(t_n))$, the limit $\lim_{n \to \infty} \sigma_n^2$. exists and is positive;
        \item $\lim_{n \to \infty} \overline f_n(t_n) \in (0, \infty)$.
    \end{enumerate}
    Then \[\overline f_n(t_n) \sqrt{n} \frac{X_{(k)} - t_n}{\sigma_n}\] converges to the standard normal in distribution.
\end{lemma}
\begin{remark} \label{remark:order}
    An important special case of the above lemma is when $F_i(t) = F(t - Y_i)$, where $F$ is a continuous and invertible c.d.f. and $(Y_n)_n$ are i.i.d. random variables. In this case, $\overline F_n$ is invertible so we may find a sequence $(t_n)_n$ such that $\overline F_n(t_n) = k/n$, which by continuity must converge to the unique $t$ which satisfies $\E F(t - Y_1) = \lambda$, so condition (ii) is satisfied. On the other hand the strong law of large numbers shows that conditions (iii) and (iv) hold almost surely.
\end{remark}

\begin{proof}[Proof of Lemma \ref{lemma:order}]
    Observe that for any $t \in \R$,
    \[\Pr(X_{(k)} \le t) = \Pr\left(\sum_{i=1}^n \1_{\{X_i \le t\}} \ge k\right)\]
    Each random variable $\1_{\{X_i \le t\}}$ is independent, and moreover their first three moments are
    \begin{align*}
        \E \1_{\{X_i \le t\}} &= F_i(t)\\
        \E (\1_{\{X_i \le t\}} - F_i(t))^2 &= F_i(t)(1-F_i(t))\\
        \E (\1_{\{X_i \le t\}} - \Pr(X_i \le t))^3 & \le F_i(t)(1-F_i(t))
    \end{align*}
    Now, fix $s \in \R$. 
    By Taylor's theorem, there exists $x$ between $t_n$ and $t_n + s / \sqrt{n}$ such that
    \[F_i\left(t_n + \frac{s}{\sqrt{n}}\right) = F_i(t_n) + f_i(x) \frac{s}{\sqrt{n}} = F_i(t_n) + f_i(t_n) O(n^{-1/2})\]
    since $s$ is constant and $f_i$ is continuous. So,
    \begin{align*}
        \frac{1}{n} \sum_{i=1}^n F_i(t_n + \frac{s}{\sqrt{n}})(1-F_i(t_n + \frac{s}{\sqrt{n}})) &= \frac{1}{n} \sum_{i=1}^n \left(F_i(t_n)(1-F_i(t_n)) + O(n^{-1/2}) f_i(t_n)\right)\\
        &= \sigma_n^2 + O(n^{-1/2})
    \end{align*}
    as $\sum_{i=1}^n f_i(t_n) = O(n)$ by assumption.
    Then
    \[\frac{\sum_{i=1}^n \E (\1_{\{X_i \le t\}} - F_i(t))^3}{\left(\sum_{i=1}^n \E (\1_{\{X_i \le t\}} - F_i(t))^2\right)^{3/2}} \le \frac{\sum_{i=1}^n F_i(t)(1-F_i(t))}{\left(\sum_{i=1}^n F_i(t)(1-F_i(t))\right)^{3/2}} = \frac{1}{\sigma_n \sqrt{n}} \to 0\]
    Hence Lyapunov's CLT condition is satisfied, and so the random variable
    \[S_t = \frac{\sum_{i=1}^n \left(\1_{\{X_i \le t\}} - F_i(t)\right)}{\sigma_n \sqrt{n}}\]
    converges to the standard normal in distribution for $t = t_n + s / \sqrt{n}$.
    Now applying the second order version of Taylor's theorem,
    \[F_i(t_n + s/\sqrt{n}) = F_i(t_n) + f_i(t) \frac{s}{\sqrt{n}} + O(n^{-1}).\]
    which gives
    \[\frac{\sum_{i=1}^n F_i(t_n + \frac{s}{\sqrt{n}}) - k}{\sqrt{n}} = \sqrt{n} \left(\frac{1}{n} \sum_{i=1}^n (F_i(t_n) + f_i(t_n) \frac{s}{\sqrt{n}} + O(n^{-1})) - \frac{k}{n}\right) = s \cdot \overline f_n(t_n) + o(1)\]
    as by assumption
    \[|\overline F_n(t_n) - k/n| = o(n^{-1/2}).\]
    Then we have
    \begin{align*}
        \frac{k - \sum_{i=1}^n F_i(t_n + \frac{s}{\sqrt{n}})}{\sqrt{\sum_{i=1}^n F_i(t_n + \frac{s}{\sqrt{n}})(1-F_i(t_n + \frac{s}{\sqrt{n}}))}} = -\frac{s}{\sigma_n} \overline f_n(t_n) + o(1).
    \end{align*}
    Combining all of the above, with $Z \sim \mathcal N(0, 1)$, we may conclude that
    \[\Pr\left(\sqrt{n} \bar f_n(t_n) \frac{X_{(k)} - t_n}{\sigma_n} \le s\right) = \Pr(S_t \ge -s + o(1)) \to \Pr(Z \le s)\]
    As this holds for all $s \in \R$, the claim follows.
\end{proof}

\begin{lemma}\label{lemma:normalapprox}
    Let $\Phi \colon \R \to [0, 1]$ be the c.d.f. of the standard normal distribution. Then for $t \ge 0$,
    \[\frac{1}{2} + \frac{t}{\sqrt{2\pi}} - O(t^3) \le \Phi(t) \le \frac{1}{2} + \frac{t}{\sqrt{2\pi}}.\]
\end{lemma}
\begin{proof}
    Let $\phi \colon \R \to \R$ be the p.d.f. of the standard normal distribution. By Taylor's theorem, there exists $0 \le s \le t$ such that $\Phi(t) = \Phi(0) + t \phi(s)$.
    By the monotonicity of $\phi$ for $x \ge 0$, it follows that
    \[\Phi(0) + t \phi(t) \le \Phi(t) \le \Phi(0) + t \phi(0).\]
    Substituting, we obtain
    \[\frac{1}{2} + \frac{t}{\sqrt{2\pi}} e^{-t^2 / 2} \le \Phi(t) \le \frac{1}{2} + \frac{t}{\sqrt{2\pi}}\]
    Using the inequality $e^{-t^2 / 2} \ge 1 - t^2 / 2$ completes the proof.
\end{proof}

\begin{lemma}\label{lemma:chernoff}
Let $X_1, \ldots, X_n$ be independent random variables taking values in $[0, 1]$ almost surely. Set $\mu = \sum_{i=1}^n \E X_i$. Then for $t > 0$,
\begin{align*}
    \Pr\left(\sum_{i=1}^n X_i \ge \mu + t\right) &\le \exp\left(- \frac{t^2}{2\mu + t}\right)\\
    \Pr\left(\sum_{i=1}^n X_i \le \mu - t\right) &\le \exp\left(- \frac{t^2}{2\mu - t}\right).
\end{align*}    
\end{lemma}
\begin{proof}
    Let $p_i = \E X_i$. The Chernoff bound gives for any $\lambda > 0$
    \[\Pr\left(\sum_{i=1}^n X_i \ge \mu + t\right) \le e^{-\lambda(\mu + t)} \E \exp\left(\lambda \sum_{i=1}^n X_i\right).\]
    By independence,
    \[\E \exp\left(\lambda \sum_{i=1}^n X_i\right) = \prod_{i=1}^n \E e^{\lambda X_i}.\]
    By convexity, $e^{\lambda X_i} \le 1 + X_i (e^\lambda - 1)$ and thus by monotonicity 
    \[\E e^{\lambda X_i} \le 1 + p_i (e^{\lambda } - 1) \le \exp\left(p_i(e^\lambda - 1)\right).\] Hence,
    \[\Pr\left(\sum_{i=1}^n X_i \ge \mu + t\right) \le \exp\left(\mu(e^\lambda - 1) - \lambda(\mu + t)\right).\]
    Taking $\lambda = \log(1 + \frac{t}{\mu})$ and using the inequality $\log(1+x) \ge \frac{2x}{2+x}$ gives
    \begin{align*}
        \Pr\left(\sum_{i=1}^n X_i \ge \mu + t\right) &\le \exp\left(t - 2\frac{t(\mu + t)}{2\mu+t} \right) = \exp\left(-\frac{t^2}{2\mu+t}\right).
    \end{align*}
    For the lower tail, for any $\lambda > 0$, 
    \[\Pr\left(\sum_{i=1}^n X_i \le \mu - t\right) \le e^{\lambda(\mu - t)} \prod_{i=1}^n \E e^{-\lambda X_i}.\]
    By convexity and monotonicity, we have
    \[\E e^{-\lambda X_i} \le 1 - p_i(1- e^{-\lambda}) \le \exp\left(-p_i(1-e^{-\lambda})\right)\]
    and so
    \[\Pr\left(\sum_{i=1}^n X_i \le \mu - t\right) \le \exp\left(\lambda(\mu - t) - \mu(1-e^{-\lambda})\right).\]
    For $\lambda = -\log(1 - \frac{t}{\mu}) \ge \frac{2t}{2\mu - t}$, we have
    \[\Pr\left(\sum_{i=1}^n X_i \le \mu - t\right) \le \exp\left(\frac{2t/\mu}{2-t/\mu}(\mu - t) - t\right) = \exp\left(-\frac{t^2}{2\mu - t}\right).\]
\end{proof}

\begin{lemma}\label{lemma:varfn}
Let $X$ be a random variable with $\var X < \infty$ and $g \colon \R \to \R$ an $L$-Lipschitz function. Then $\var g(X) \le L^2 \var X$.
\end{lemma}
\begin{proof}
    For any random variable $Y$, we have
    \[\var Y = \E Y^2 - (\E Y)^2 \le \E Y^2\]
    Hence,
    \[\var g(X) = \var (g(X) - g(\E X)) \le \E(g(X) - g(\E X))^2.\]
    Since $g$ is $L$-Lipschitz, we have $(g(X) - g(\E X))^2 \le L^2 \cdot (X - \E X)^2$ and so by monotonicity
    \[\var g(X) \le L^2 \E(X - \E X)^2 = L^2 \var X. \]
\end{proof}

\section{Discussion}
We have described a model of the brain at the neuronal level in which the statistical structure of stimuli can be recorded and reproduced through assemblies and sampling. Key ingredients of this model, inherited from NEMO, are brain areas, random synapses, plasticity, and inhibition-induced competition; to these we add here random noise, and an additive plasticity rule enabling statistical learning, which may be of interest on its own.  We provide both simulations demonstrating statistical learning phenomena, as well as proofs of a theoretically tractable special case. It would be interesting to extend these results to more complex graphical models in efficient ways. Although we have focused on Markov chains, our methods can be used to show that via the presentation of an appropriate sequence of stimuli, NEMO can perform arbitrary probabilistic computation in conjunction with the Turing machine simulation in NEMO previously shown \cite{dabagia2024computation}. 

A particularly interesting observation is the emergence of the softmax, or Boltzmann distribution, as the probability that an assembly fires as a function of its weight, the first result of its kind to our knowledge in the study of the brain. Our model also makes a novel prediction about the precise Hebbian plasticity rule used by the brain necessary to support this form of statistical learning. 

A fertile source of inspiration for this work is naturally language acquisition (surely one of the human brain's most remarkable abilities), not least due to the substantial amount of experimental work in psychology dedicated to understanding the statistical dimensions of this task. We imagine the statistical sensitivity our model supports the most rudimentary phases of language learning; for instance, reproducing the phonological structure of words while babbling \cite{boyssonbardies1991adaptation}, or discovering word boundaries \cite{saffran1996statistical}. Our simple language experiments are in line with this perspective, demonstrating a ``babbler'' in NEMO which generates locally-plausible word sequences but lacks higher-level structure. In Kahneman's two-systems theory \cite{kahneman2011thinking}, this form of statistical learning might be regarded as the ``fast'' system, which requires interaction with the symbolic reasoning of the ``slow'' system to produce grammatical language. This will also surely be a rich direction for future work --- in particular, the present model does not assume nor seem to discover any structure in its input beyond local co-occurrence statistics. One can imagine more complex models which learn to construct syntax trees of their input (perhaps again using statistical cues), and sampling could then occur at higher levels in the tree. This parallels the observation in language acquisition that the phase wherein the brain ''commits itself'' to the particular patterns of its native language seems to be crucial to a speaker's development \cite{kuhl2004early}.

\begin{ack}
MD and SV are supported by NSF award CCF-2106444, an NSF GRFP fellowship, and a Simons Investigator award. DM and CP are supported by NSF awards 2229929 and 2212233, and by a grant from Softbank.
\end{ack}

{
\small
\printbibliography

}





\end{document}